\documentclass[11pt]{article}
\usepackage{amsmath}
\usepackage{amsfonts}
\usepackage{amssymb}
\begin{document}
\newcommand{\half}{\frac{1}{2}}
\newcommand{\osixth}{\frac{1}{6}}
\newcommand\sech{{\rm sech}}
\newcommand{\IZ}[1]{\bar{#1}}
\newcommand{\Deta}{\eta^{\dagger}}
\newcommand\Bpsi{\boldsymbol{\psi}}
\newtheorem{theorem}{Theorem}
\newtheorem{lemma}{Lemma}
\newtheorem{definition}{Definition}
\newtheorem{corollary}{Corollary}
\newtheorem{remark}{Remark}
\title{
\begin{flushright}
  \small UMP-98/?
\end{flushright}
\vskip1.0cm
\large\bf
The Lanczos Algorithm for extensive Many-Body Systems in the 
Thermodynamic Limit}
\author
{\large N.S. Witte\footnotemark[1],\\
{\it Research Centre for High Energy Physics,}\\
{\it School of Physics, University of Melbourne,}\\
{\it Parkville, Victoria 3052, AUSTRALIA.}\\
{\large D. Bessis}\\
{\it Service de Physique Th\'eorique,}\\
{\it Centre d'Etudes Nucl\'eaires de Saclay,}\\
{\it F-91191 Gif-Sur-Yvette Cedex}\\
{\it FRANCE} }
\maketitle
\begin{abstract}
We establish rigourously the scaling properties of the Lanczos process
applied to an arbitrary extensive Many-Body System which is carried to 
convergence $ n \to \infty $ and the thermodynamic limit $ N \to \infty $ 
taken. 
In this limit the solution for the limiting Lanczos coefficients
are found exactly and generally through two equivalent sets of 
equations, given initial knowledge of the exact cumulant generating
function.
The measure and the Orthogonal Polynomial System associated with the
Lanczos process in this regime are also given explicitly.
Some important representations of these Lanczos functions are
given, including Taylor series expansions, and theorems controlling their 
general properties are proven. 

\end{abstract}
\renewcommand{\thefootnote}{\fnsymbol{footnote}}
\footnotetext[1]{E-mail: {\tt nsw@physics.unimelb.edu.au}}
\renewcommand{\thefootnote}{\arabic{footnote}}
\vfill\eject
\setcounter{section}{0}
\stepcounter{section}
\section*{\Roman{section}. Introduction}

The Lanczos Algorithm is one of the few reliable and general methods for 
computing the ground state and excited state properties of strongly interacting 
quantum Many-body Systems. It has been traditionally employed as a numerical
technique on small finite systems, with attendant round-off error
problems, although the main obstacle to its further development has been 
the rapid growth of the number of basis states with system size.
The reader is referred to a review of the applications of this 
method\cite{lanczos-D-94} in strongly correlated electron problems.
In this work we examine the Lanczos process in the context of the extensive 
quantum Many-Body Systems, where it is employed entirely in an exact manner 
and where the thermodynamic limit is taken.
So in complete contrast to the traditional use of the Lanczos algorithm - we 
completely circumvent the issues of loss of orthogonality due to round-off 
errors and the inability to approach the thermodynamic limit because of
the requirement to construct a full basis on the cluster.
The systems we have in mind are those with an infinite number of degrees
of freedom, yet are extensive, in that all total averages of any 
physical quantity scale linearly with the numbers of degrees of freedom
however quantified. These would include all condensed matter systems
with sufficiently local interactions (the precise conditions need to be
clarified, but it is clear which specific systems obey extensivity) and
Quantum Field Theories, with the proviso that the spectrum is bounded
below (in some cases there is also an upper bound too).

After noting some of the advantageous features of the algorithm in general we 
discuss the scaling behaviour of the Lanczos Process as it approaches 
convergence and as the thermodynamic limit is taken. Central to this approach
is the manifestation of extensivity through a description based on the Cumulant
Generating Function, which we take to be given. We then derive a set of
general integral equations which define the scaled Lanczos functions in the 
thermodynamic limit, which can be explicitly and exactly solved for certain 
integrable models, or employed in a truncated manner for non-integrable models.
An alternative formulation is also given which expresses the equivalence of
the Lanczos Process with the continuum Toda Lattice Model treated as a 
boundary value problem. Finally we state some general results concerning the 
behaviour of the Lanczos functions.

\vfill\eject

\stepcounter{section}
\section*{\Roman{section}. The Lanczos Process, Orthogonal Polynomials and Moments}

The Lanczos Algorithm or Process\cite{lanczos,eigen-P,eigen-C} begins with 
a trial state $ |\psi_0 \rangle $ appropriate to the model and the symmetries 
of the phase being investigated. From this the Lanczos recurrence generates a 
sequence of orthonormal states $ \{ | \psi_n \rangle \}^{\infty}_{n=1} $ and 
Lanczos coefficients $ \{ \alpha_n(N) \}^{\infty}_{n=0} $ and 
$ \{ \beta_n(N) \}^{\infty}_{n=1} $, thus
\begin{equation}
  \hat{H} |\psi_n \rangle =
  \beta_n |\psi_{n-1} \rangle + \alpha_n |\psi_n \rangle
 + \beta_{n+1} |\psi_{n+1} \rangle \ ,
\label{lanczos-recur}
\end{equation}
with the Lanczos coefficients being defined
\begin{equation}
 \begin{split}
 \alpha_{n}
  & = \langle \psi_{n}| \hat{H} |\psi_{n}\rangle \ , \\
 \beta_{n}
  & = \langle \psi_{n-1}| \hat{H} |\psi_{n}\rangle \ . \\
\end{split}
\label{lanczos-coeff}
\end{equation}
We distinguish a total or extensive operator or variable such as $ H $
from its density or intensive counterpart by $ \hat{H} $.
In this basis the transformed Hamiltonian takes the following tridiagonal form
\begin{equation}
  T_{n} =
  \left(
   \begin{array}{cccccc}
    \alpha_0 & \beta_1  &         &             &              &          \\
    \beta_1  & \alpha_1 & \beta_2 &             &              &          \\
             &          & \ddots  &             &              &          \\
             &          &         & \beta_{n-1} & \alpha_{n-1} & \beta_n  \\
             &          &         &             & \beta_n      & \alpha_n  
   \end{array}
  \right) \ .
\label{tridiag}
\end{equation}
As such the Lanczos process is one of the Krylov subspace 
methods\cite{eigen-S}, in that at a finite step $ n $, the eigenvectors
belong to the Krylov Subspace 
$ {\rm Span}\{ |\psi_0 \rangle, \hat{H} |\psi_0 \rangle, 
           \hat{H}^2 |\psi_0 \rangle, \ldots \hat{H}^n |\psi_0 \rangle \} $.

In the Many-Body context one would iterate the Lanczos Process until 
termination whereupon the Hilbert space is exhausted (at this point one of the 
$ \beta_{n_T} = 0 $, where $ n_T $ is the dimension of 
the Hilbert space in the sector defined by the ground state), or until the 
process has converged according to some arbitrary criteria $ n \to n_C $. 
Then one would perform the thermodynamic limit $ N \to \infty $
where it should be understood that the above conclusion of the Lanczos
process is also dependent on the system size, that is to say $ n_T(N), n_C(N) $.
These cutoffs are monotonically increasing functions of the system size so they 
will all tend to $ \infty $ in the thermodynamic limit as well.
Taking the limits in the reverse order clearly leads to nonsensical results, 
as taking $ N \to \infty $ with $ n $ fixed produces $ \alpha_n \to c_1 $ and 
$ \beta_n \to 0 $. The great virtue of the Lanczos process is that it can be 
shown to converge essentially exponentially fast with respect to iteration 
number, using the Kaniel-Paige-Saad exact bounds\cite{leb-K,leb-P,leb-S} for 
the rate of convergence. This means that convergence occurs within a very small 
subspace of the total Hilbert space, so that $ n_C << n_T $.

The Lanczos process is entirely equivalent to the 3-term recurrence for an 
Orthogonal Polynomial System\cite{ops-S,ops-C,ops-F}, however we consider a 
slight generalisation of the preceding process to one with a single parameter
evolution (a "time" $ t $). In this construction we are continuing a 
development begun by Lindsay\cite{mm-L89a} and Chen and 
Ismail\cite{matrix-CI-97}, which will lead to some powerful tools in treating 
the Lanczos process. The measure, or that component which is absolutely 
continuous, is defined by the weight function
\begin{equation}
   w(\epsilon,t) = e^{-u(\epsilon)+tN\epsilon} \ ,
\label{t-weight}
\end{equation}
on the real line $ \epsilon \in {\mathbb R} $. Our system under study is 
described by the initial value of the system at $ t = 0 $ and often we will 
suppress this argument for the sake of simplicity. This measure defines a 
system of monic Orthogonal Polynomials $ \{P_n(\epsilon,t) \}^{\infty}_{n=0} $ 
with an orthogonality relation
\begin{equation}
 \int^{+\infty}_{-\infty} d\epsilon\; w(\epsilon,t) 
      P_m(\epsilon,t) P_n(\epsilon,t) = h_n(t) \delta_{mn} \ ,
\label{orthog}
\end{equation}
and normalisation $ h_{n}(t) $. This is equivalent to the following three-term 
recurrence relation
\begin{equation}
   P_{n+1}(\epsilon,t) = (\epsilon-\alpha_n(t)) P_{n}(\epsilon,t)
                               - \beta^2_n(t) P_{n-1}(\epsilon,t) \ ,
\label{ops-recur}
\end{equation}
with the recursion coefficients $ \alpha_{n}(t) $ real for $ n \geq 0 $ and
$ \beta^2_{n}(t) $ real and positive for $ n > 0 $. By convention we take 
$ \beta^2_{0} = 1 $. It can be readily shown that the Lanczos coefficients are 
given in terms of the normalisation thus
\begin{equation}
\begin{split}
   \alpha_{n}(t) & = {1\over Nh_{n}(t)}{d \over dt}h_{n}(t)  \ ,
   \\  
   \beta^2_{n}(t) & = {h_{n}(t) \over h_{n-1}(t)} \ .
\end{split} 
\label{ops-coeff}
\end{equation}
The direct connection between the Lanczos Process and the OPS are given by the 
determinant relation of the characteristic polynomial 
\begin{equation}
   P_{n+1}(\epsilon) = (-)^{n+1}|T_{n}-\epsilon I_{n+1}| \ ,
\label{tridiag-eigen}
\end{equation}
so that the zeros of the Orthogonal Polynomial are eigenvalues of Hamiltonian.

Some comments are in order regarding the differences, or more accurately
the special character, of these Orthogonal Polynomials with respect to
the generic OPS or with some of the scaling versions of OPS\cite{ops-vA-90}.
These OPS have been termed Many-Body OPS, but could be equally described as
extensive OPS. They all have an additional, essential parameter to the
generic OPS, the system size $ N $, which appears in both the gross scaling 
factors (the `external' scaling such as in the energy densities $ \epsilon $ 
defined by $ E = N\epsilon $), but also internally in the 3-term recurrence 
coefficients, in the Polynomials themselves and in other derived quantities. 
The internal dependence in the Lanczos coefficients on the system size is not 
at all apparent and the most transparent way that extensive scaling properties 
can be exhibited is through the Cumulant Generating Function, (CGF), which 
hitherto has played no role in Orthogonal Polynomial Theory. In fact the CGF 
is central to this class of OPS rather than the moments, and is in a practical 
sense the starting point in any application of the Formalism to physical Models.
For all models it is clear that the ground state energy $ E_0 $ is proportional 
to $ N $ and unbounded in the thermodynamic limit, and similarly the total 
Lanczos coefficients (as opposed to the densities) are unbounded as 
$ n \to \infty $ for fixed $ N $. When everything is recast in terms of 
densities the spectrum is bounded below by $ \epsilon_0 $ and in many models 
will also be bounded above, and similarly the density Lanczos coefficients are 
bounded.  Another difference that Many-Body OPS exhibit in comparison to 
general OPS is, as we have noted above, the three-term recurrence will 
terminate exactly at $ n = n_{T} $, although this will never present any 
problems as this is exponentially large.

The Lanczos process is intimately connected with the Hamburger moment 
problem\cite{moment-ST,moment-A}, via the Resolvent operator
\begin{equation}
     R(\epsilon) = \langle {1 \over \epsilon-\hat{H}} \rangle 
     \qquad \epsilon \notin {\rm Supp}[d\rho] \ .
\label{resolvent-defn}
\end{equation}
Its formal Laurent series establishes a direct link with Hamiltonian moments
\begin{equation}
     R(\epsilon) = \sum^{\infty}_{i=0} {\mu_{i} \over \epsilon^{i+1}} \ ,
\label{resolvent-moment}
\end{equation}
where these moments are defined as expectation values with respect to the trial 
state referred to above
\begin{equation}
        \mu_n \equiv \langle \hat{H}^n \rangle\ , \qquad \mu_0 = 1 \ .
\label{H-moment}
\end{equation}
The resolvent has a real Jacobi-fraction continued fraction 
representation\cite{cf-JT,cf-LW}
\begin{equation}
  R(\epsilon) =
  - {\Large\mathbf{K}}^{\infty}_{n=0}
              -\left({ \beta^2_n \over \epsilon-\alpha_n }\right) \ ,
 \label{resolvent-J-cf}
\end{equation}
with elements coming from the Lanczos coefficients.

An equivalent description to that of the Hamiltonian moments is to formulate 
everything in terms of cumulants or connected moments\cite{moment-K,cumulant-K}
$ \{ \nu_n \}^{\infty}_{n=1} $, and to ignore all corrections which vanish in 
the thermodynamic limit $ N \to \infty $. Cumulants scale directly with the 
size of the system so that for the extensive Many-Body Problem we have
\begin{equation}
  \nu_n = c_n N + {\rm o}(1)
\label{cum-GS}
\end{equation}
in the ground state sector, or 
\begin{equation}
  \nu_n = c_n N + m_n + {\rm o}(1)
\label{cum-ES}
\end{equation}
in any other sector\cite{pexp-tgap-HWW}. This also means that no finite-size 
scaling can be performed given that only the limiting quantities are retained 
here and boundary condition effects do not appear. The foundation ingredient 
is the Moment Generating Function which is related to the Cumulant Generating 
Function in the following way.
\begin{definition}
The Moment Generating Function (MGF) $ M(t) $ and the Cumulant Generating 
Functions (CGF) $ F(t) $ are defined by,
\begin{equation}
    M(t) \equiv \langle e^{tH} \rangle
         = \sum^{\infty}_{n=0} \mu_{n} {t^n \over n!}
         = \exp\left( \sum^{\infty}_{n=1} \nu_{n} {t^n \over n!} \right)
         \equiv \exp(NF(t)) \ .
\label{gen-funs}
\end{equation}
\end{definition}
Some examples of Cumulant Generating Functions include the isotropic XY
model using the z-polarised N\'eel state as the trial state\cite{texp-W-97}
\begin{equation}
  F(t) = {1 \over \pi}
           \int^{\pi/2}_{0}\, dq \log\cosh(t\cos q) \ ,
\label{xy-cgf}
\end{equation}
and the Ising model in a transverse field using the disordered state as the
trial state, and coupling constant $ x $\cite{alm-itf-W-98}
\begin{equation}
  F(t) =
    {1 \over 2\pi} \int^{\pi}_{0} dq\,
         \ln \left[ \cosh(2t\epsilon_{q}) 
                  - {(\cos q\!+\!x) \over \epsilon_{q}}\sinh(2t\epsilon_{q})
            \right] \ ,
\label{itf-cgf}
\end{equation}
where the quasiparticle energies $ \epsilon_{k} $ are defined by
$ \epsilon^2_{q} = 1+x^2+2x\cos q $.

\begin{definition}
The Determinants of the Moment Matrices $ \Delta_{n}(t) $ for $ n \geq 0 $ 
are defined by the Hankel form -
\begin{equation}
   \Delta_n(t)
   = |M^{(i+j-2)}(t)|^{n+1}_{i,j=1} \ .
\label{hankel-tD}
\end{equation}
\end{definition}
The direct relationship from moments to the Lanczos coefficients which is 
established in this way is via the construction of a sequence of Hankel 
determinants of the Moment Matrices and their Selberg-type integral 
representation\cite{ops-S}
\begin{equation}
   \Delta_n(t)
   = {1\over (n\!+\!1)!} \int^{+\infty}_{-\infty} 
        \prod^{n+1}_{k=1} d\epsilon_k\; w(\epsilon_k,t)
        \prod_{1\leq i<j \leq n+1} |\epsilon_i-\epsilon_j|^2
             \ .
\label{hankel-tS}
\end{equation}
These determinants are related to the normalisations via
\begin{equation}
   \Delta_n(t) = \prod_{j \leq n} h_j(t) \ .
\label{hankel-norm}
\end{equation}
\begin{definition}
Our final definition, that of the Lanczos $ L $-function, is
\begin{equation}
   N^2 L_{n}(t) = 
   { \Delta_{n}(t) \Delta_{n-2}(t)
                   \over \Delta^2_{n-1}(t) } \ ,
\label{l-function}
\end{equation}
for $ n \geq 1 $ and $ L_{0}(t) = M(t) $. 
\end{definition}
The converse result is then
\begin{equation}
    \Delta_{n}(t) = N^{n(n+1)} \prod^{n}_{k=0} L^{n+1-k}_{k}(t) \ ,
\label{D-function}
\end{equation}
for $ n \geq 1 $. From these the Lanczos coefficients are given simply by
\begin{equation}
\begin{split}
   \alpha_n(t)  & = 
   {1\over N} \sum^{n}_{j=0}
           { L'_{j}(t) \over L_{j}(t) } \ ,
             \\
   \beta^2_n(t)  & = L_{n}(t)
             \ .
\end{split}
\label{coeff-l-function}
\end{equation}

\begin{theorem}
The equation of motion for the Lanczos $ L $-functions is
\begin{equation}
   L_{n}(t) = {1\over N} 
   \sum^{n}_{j=1} {j \over N} D_{t}^2 \log L_{n-j}(t) \ .
\label{l-function-recur}
\end{equation}
with the initial condition on the recurrence given by $ \log L_{0}(t) = NF(t) $ 
for all $ t $.
\end{theorem}
The advantage of introducing evolution into the Lanczos Process is that 
Sylvester's Theorem applied to the Hankel
determinants\cite{positive-K},
\begin{equation}
   \Delta_{n+1}(t) \Delta_{n-1}(t) =
   \Delta_{n}(t) \Delta''_{n}(t) - \left( \Delta'_{n}(t) \right)^2 \,
\label{sylvester}
\end{equation}
so that the theorem follows directly from this.$ \square $

The first few members of the Lanczos $ L $-sequence are
\begin{equation}
\begin{split}
  L_{1}(t) 
  & = {1\over N} F''(t) \ ,
  \\
  L_{2}(t)
  & = {2\over N} F''(t) + 
      {1\over N^2}{ F^{(2)}F^{(4)}-(F^{(3)})^2 \over (F^{(2)})^2 }
  \ .
\end{split}
\label{l-function-first}
\end{equation}

The consequence of Sylvester's theorem for the evolution of the $ \Delta_{n} $ 
is the following theorem
\begin{theorem}
The $ \Delta_{n}(t) $ obey the following differential-difference equation
\begin{equation}
  \exp\left\{ \log\Delta_{n+1} + \log\Delta_{n-1} - 2 \log\Delta_{n}
      \right\}
  = D^2_{t} \log\Delta_{n} \ ,
\label{D-function-eqn}
\end{equation}
with the boundary value $ \log\Delta_{0} = NF(t) $ and conventionally
$ \Delta_{-1} = 1 $.
\end{theorem}
This follows directly from Sylvester's Identity.$ \square $

This evolution equation is just the finite Toda Lattice equation of 
motion\cite{ops-T}, and this point has been previously noted in
Ref.~\cite{matrix-CI-97}.

\vfill\eject

\stepcounter{section}
\section*{\Roman{section}. Scaling in the Thermodynamic Limit}

As was discussed earlier there are two limiting processes that one must 
consider when the thermodynamic limit is taken in the Lanczos Algorithm, both 
$ n, N \to \infty $, and the issue then is what mutual relationship exists 
between them in the limit. One can view this limiting process in the $ 1/n $ 
vs. $ 1/N $ plane and then consider along what types of paths must one approach 
the origin. We shall find that the general relationship is $ n, N \to \infty $ 
with $ s \equiv n/N $ fixed, although for systems at criticality it seems 
inevitable that $ s $ will become unbounded in the analysis.
A consequence of these ideas is the confluence property of the Lanczos 
coefficients as $ n,N \to \infty $ at fixed $ s=n/N $
\begin{equation}
\begin{split}
   \alpha_n(N)  & = \alpha(s) + {\rm O}(1/N) \ , \\
   \beta^2_n(N) & = \beta^2(s) + {\rm O}(1/N) \ . \\
\end{split}
\label{lanczos-confl}
\end{equation}
There are a number of ways to see this approach to the thermodynamic limit.

Using the explicit forms connecting cumulants and moments, and a direct 
evaluation of the Hankel determinants one can prove\cite{pexp-proof-WH} for 
general $ n $ and $ N $ that the Lanczos coefficients have a leading order 
scaling in $ s=n/N $ for the first two orders of an expansion in large $ N $.
Actually this expansion is valid for all $ n $ not just for large values and 
thus includes all the subdominant contributions. Thus
\begin{equation}
\begin{split}
   \alpha_{n} = 
 & c_1 N  + n \left[ {c_{3}\over c_{2}} \right]
 \\
 & \qquad + {1\over 2}n(n-1)
    \left[ {3 c_3^3\!-\!4 c_2 c_3 c_4\!+\!c_2^2 c_5 \over 2 c_2^4} \right]
               {1\over N}
 \\
 & \qquad\qquad + \dots \ ,
\end{split}
\label{alpha-exp}
\end{equation}
for $ n \ge 0 $, and
\begin{equation}
\begin{split}
   \beta_{n}^{2}
 & = n c_{2} N
          + {1\over 2}n(n-1)
          \left[ {c_{2}c_{4}\!-\!c_{3}^{2}\over c_{2}^{2}} \right]
 \\
 & \quad + {1\over 6}n(n-1)(n-2)
     \left[ {-12 c_{3}^4\!+\!21 c_{2} c_{3}^2 c_{4}
              \!-\!4 c_{2}^2 c_{4}^2\!-\!6 c_{2}^2 c_{3} c_{5}
                \!+\!c_{2}^3 c_{6}
                             \over 2 c_{2}^5} \right] {1\over N}
 \\
 & \qquad  + \dots \ ,
\end{split}
\label{beta-exp}
\end{equation}
for $ n \ge 1 $.
However this approach cannot be generalised to higher orders and therefore for 
the full exact Lanczos coefficients. The first two terms in the above 
expansions were also proven by Lindsay using the Sylvester Identity in the 
statistical context\cite{mm-L89a} but no further, while this form for the 
higher terms (but finite numbers) was conjectured in Reference\cite{pexp-1st-H}.
We shall find that use of the Sylvester Identity allows one to very easily 
recover this result, to in fact go to much higher orders in constructing 
explicit forms and to prove this type of scaling in a completely general way.

\begin{lemma}
The Lanczos $ L $-function $ L_{n}(t,N) $ is a rational function of $ 1/N $ 
for fixed $ n $, and all $ t $.
\end{lemma}
The Difference-Differential Eq.~(\ref{l-function-recur}) is of finite order in 
$ j/N $ and $ t $, so the result follows.$ \square $

Also for fixed $ n $ we have
\begin{equation}
  \lim_{N \to \infty} L_{n}(t,N) = 0 \ ,
\label{l-function-limit}
\end{equation}
and specifically the leading order term is $ {\rm O}(N^{-1}) $ which arises 
from the $ j = n $ term in the sum. Therefore we can expand this function in a 
descending series in $ N^{-1} $, thus
\begin{equation}
   L_{n}(t,N) = \sum_{p \geq 1} 
     { l_{np}(t) \over N^{p} } \ ,
\label{l-function-exp}
\end{equation}
and defining the connected series related by
\begin{equation}
    \sum_{p \geq 1} { m_{np}(t) \over N^{p} }
  \equiv \log\left( 1+\sum_{p \geq 1} { l_{np+1}/l_{n1} \over N^{p} }
        \right) \ .
\label{l-function-coeff-1}
\end{equation}
This last relation can be rendered into an explicit form
\begin{equation}
  m_{np} = - \sum_{\sum_{i} q_{i}r_{i} = p} (\sum_{i} q_{i}\!-\!1)! 
         \prod_{i} {1\over q_{i}!} 
         \left({-l_{nr_{i}+1} \over l_{n1}} \right)^{q_{i}} \ .
\label{l-function-coeff-2}
\end{equation}
It is actually necessary to perform an expansion of this type because it 
combines the iteration number ($ n $) dependence of the numerator and 
denominator which are both essential in the following results.

Then one can establish a hierarchy of equations for these coefficients
\begin{equation}
\begin{split}
  l_{n1}(t) 
  & = n F''(t) \ ,
  \\
  l_{n2}(t)
  & = \sum^{n-1}_{j=1} j D^2_{t} \log l_{n-j1}(t) \ ,
  \\
  l_{np}(t) 
  & = \sum^{n-1}_{j=1} j m''_{n-jp-2}(t)
  \qquad \text{for } p \geq 3 \ ,
\end{split}
\label{l-function-exp-recur}
\end{equation}
for $ n \geq 1 $ whilst for $ n=0 $ we have $ l_{np}(t) = 0 $ as 
$ L_{0}(t,N) = \exp( NF(t) ) $. The first members of this hierarchy can be 
easily solved for yielding
\begin{equation}
\begin{split}
  l_{n1}(t) 
  & = n F''(t) \ ,
  \\
  l_{n2}(t)
  & = \half n(n-1) { F^{(2)}F^{(4)} - (F^{(3)})^2 \over (F^{(2)})^2 } \ ,
  \\
  l_{n3}(t)
  & = \frac{1}{12} n(n-1)(n-2) 
      \left( { F^{(2)}F^{(4)} - (F^{(3)})^2 \over (F^{(2)})^3 }
      \right)^{(2)} \ ,
\end{split}
\label{l-function-exp-first}
\end{equation}
and from these it is easy to establish the leading order terms already found 
in Eq.~(\ref{alpha-exp},\ref{beta-exp}).

\begin{lemma}
The hierarchy coefficients $ l_{np}(t), m_{np}(t) $ are polynomials in $ n $.
\end{lemma}
These coefficients are constructed from a finite difference equation in $ n $.
$ \square $

\begin{theorem}
The hierarchy coefficients $ l_{np}(t), m_{np}(t) $ are polynomials of degree 
$ p $ in $ n $.
\end{theorem}
This is proved by induction on $ p $ using the hierarchy equations. If we take 
$ l_{jq}(t) $ to be of degree $ q \leq p-2 $ in $ n $ then similarly for 
$ m_{jq}(t) $ and $ m''_{jq}(t) $. Now for any polynomial $ P(n) $ of degree 
$ p-2 $ in its argument then
\begin{equation}
   \sum^{n-1}_{j=1} j P(n\!-\!j) \ ,
\label{induction}
\end{equation}
is a $ p $th degree polynomial. Thus the recurrence, 
Eq.~(\ref{l-function-exp-recur}), establishes that $ l_{n+1p} $ is also a 
$ p $th degree polynomial.$ \square $

From this result it is clear that the limiting forms of the Lanczos coefficients
$ \alpha_{n}(N), \beta^2_{n}(N) $ exist when $ n, N \to \infty $ with $ n/N $ 
fixed. If the ratio is not kept constant in this limiting operation, say with 
$ n = {\rm o}(N) $ then the Lanczos coefficients will vanish in the limit, 
while if the reverse is true $ N = {\rm o}(n) $ then there will be divergent 
terms in the limit.

Given that the scaling Lanczos coefficients have been established then all the 
exact theorems for the ground state 
properties\cite{pexp-exactgse-HW,pexp-2dhm-WHW} that were predicated on this 
result now are established. The first example of these theorems was the one for 
the ground state Energy Density,
\begin{equation}
  \epsilon_0 = \inf_{s \in {\mathbb R^+}}[\alpha(s)-2\beta(s)] \ ,
\label{GSE-bot}
\end{equation}
which also has an analogue for the top of the spectrum, if this exists
\begin{equation}
  \epsilon_{\infty} = \sup_{s \in {\mathbb R^+}}[\alpha(s)+2\beta(s)] \ .
\label{GSE-top}
\end{equation}
For many models these Lanczos Functions will be bounded on the positive real 
axis, and have limits as $ s \to \infty $ on the real line. So there is a 
superficial similarity to classes of Orthogonal Polynomials whose 3-term 
recurrence coefficients have limiting values, such as the S-class, the M-class, 
or the $ M(a,b) $ classes\cite{ops-vA-90}.

\vfill\eject

\stepcounter{section}
\section*{\Roman{section}. The extensive Measure}

It is necessary to determine the OPS measure, its weight function
$ w(\epsilon) $, and this is not generally known at the outset, but rather the 
Cumulant Generating Function is. In fact it seems to be the case that the 
measures are not exactly expressible in simple terms, but the CGF or 
characteristic functions are. There is of course a direct route from a model 
system and a trial state to the Lanczos coefficients, but from many points of 
view including practical considerations the route beginning with a cumulant 
description is more useful.

\begin{theorem}
Given that the cumulant generating function $ F(-t) $ is analytic for 
$ \Re(t) > 0 $ and in the neighbourhood of the origin $ t = 0 $ the OPS weight 
function $ w(\epsilon) $ has the following asymptotic development in the 
thermodynamic limit $ N \to \infty $,
\begin{equation}
  w(\epsilon) =
  \sqrt{ N \over 2\pi F^{(2)}(\xi)}
   e^{N\left[ -\epsilon\xi + F(\xi) \right]}
    + {\rm O}(N^{-1/2}) \ ,
\label{weight-tl}
\end{equation}
where the function $ \xi(\epsilon) $ is defined implicitly by
\begin{equation}
   \epsilon = F'(\xi) \ .
\label{stationary}
\end{equation}
\end{theorem}
Starting with the definition of the cumulant generating function $ F(t) $
\begin{equation}
   \langle e^{tH} \rangle \equiv \exp\{NF(t)\}
   = \exp\left\{ N\sum_{n=1}^{\infty} {c_n \over n!} t^n \right\} \ .
\label{cgf}
\end{equation}
We assume here that this infinite series is not just formal but actually exists,
that is it has a finite radius of convergence in addition to its analytic 
character for $ \Re(t) < 0 $. However the Moment Generating Function is simply 
the analytic continuation of the characteristic function and this continuation 
is possible given its analyticity, so that a Fourier inversion of this will 
yield the weight function, 
\begin{equation}
 \begin{split}
  w(\epsilon) 
  & = {N \over 2\pi} \int^{i\gamma+\infty}_{i\gamma-\infty}
   dt\; e^{N[ -it\epsilon + F(it) ]} \ ,
  \\
  & = {N \over 2\pi i} \int^{\gamma+i\infty}_{\gamma-i\infty}
   dt\; e^{N[ t\epsilon + F(-t) ]} \qquad \Re(\gamma) > 0 \quad \ .
 \end{split}
\label{fi-character-fn}
\end{equation}
One does not require the exact inversion but only the leading order in $ N $ 
in a steepest descent approximation. In an asymptotic analysis the relevant 
function is 
\begin{equation}
   g(t) = t\epsilon + F(-t) \ ,
\label{exponent}
\end{equation}
which is analytic for all $ \Re(t) > 0 $. We will assume the existence of a 
stationary point which occurs at $ t_0 $
\begin{equation}
  \epsilon = F'(-t_0) \ ,
\label{stationary-t}
\end{equation}
and is assumed to be unique. This point is evidently real because the energy 
density is real and the CGF is a real function of a real argument (here we
define $ \xi = -t_0 $ for convenience). One requires the inversion of this 
relation for $ \xi(\epsilon) $ and this is guaranteed by the Implicit Function 
Theorem because $ F^{(2)}(\xi) > 0 $. This latter condition also implies that 
the saddle point is of order unity. Indeed one clearly has the case of 
$ F^{(2)}(t) > 0 $ for real values of $ t $ in the neighbourhood of the saddle 
point and $ F^{(2)}(t) < 0 $ for imaginary values of $ t $ in the same 
neighbourhood. Thus the path of steepest descent through the saddle point is 
parallel to the imaginary axis.  One can then apply the standard saddle point 
analysis, see Wong\cite{asymptotic-W} Section II.4, to arrive at the stated 
result. $ \square $

The corresponding example of the saddle point equation for the isotropic XY 
model is
\begin{equation}          
  \epsilon = {1\over \pi} \int^{\pi/2}_0 dq\, \cos q \tanh(\xi\cos q) \ ,
\label{xy-saddle}                           
\end{equation}                                                  
and that for the Ising model in a transverse field is
\begin{equation}
   \epsilon = {1 \over \pi}
   \int^{\pi}_{0}dq\, \epsilon_{q}
   { {\displaystyle x\!+\!\cos q \over \displaystyle \epsilon_{q}}
     + \tanh(2\xi\epsilon_{q})
     \over
     1 + {\displaystyle x\!+\!\cos q \over \displaystyle \epsilon_{q}}
         \tanh(2\xi\epsilon_{q}) } \ .
\label{itf-saddle}
\end{equation}

The first of the more obvious properties concerns the convexity of the measure 
arising in the thermodynamic limit,
\begin{theorem}
The leading order of the negative logarithm of the weight function 
$ u(\epsilon) $ is convex for all real energies $ \epsilon $.
\end{theorem}
This follows from the relationship of $ u(\epsilon) $ to the stationary point 
\begin{equation}
   {d \over d\epsilon} u(\epsilon) = N \xi(\epsilon) \ ,
\label{deriv-u}
\end{equation}
and the definition
\begin{equation}
  \epsilon = F'(\xi) \ .
\label{stationary-2}
\end{equation}
Now it can be easily seen that $ F''(t) > 0 $  for $ t $ real and the Hermitian 
Hamiltonian using the definition of $ F(t) $ in terms of the expectation value
$ NF(t) = \ln \langle \exp(tH) \rangle $.$ \square $

Some detailed, yet general information, concerning the extensive measure in the 
neighbourhood of the ground state is available. This arises from consideration 
of the overlap of the trial state with the true ground 
state\cite{cmx-overlap-C}, and its relation to the Horn-Weinstein function 
$ E(t) \equiv F'(-t) $ via 
\begin{equation}
  \left| \langle \Psi_{GS} | \psi_{0} \rangle \right|^{2} 
  = \exp\left\{ -N \int^{\infty}_{0} dt\; 
                 [ E(t)-E(\infty) ] \right\} \ .
\label{overlap}
\end{equation}
In general the limit $ E(t) $ as $ t \to \infty $ will exist, and is the ground 
state energy, and so the asymptotic properties of $ E(t) $ for $ \Re(t) > 0 $ 
as this tends to infinity is a means of classifying systems. This equivalent to 
the asymptotic properties of $ \epsilon(\xi) - \epsilon(-\infty) $ as 
$ \xi \to -\infty $ (we denote the Ground State Energy by $ \epsilon_0 $,
which is also the same as $ \epsilon(-\infty) $). In general the overlap is 
non-zero, so that $ E(t)-E(\infty) \in L^{1}[0,\infty) $ but it is possible at 
isolated points that this is not true (critical points in the model for example)
and the overlap may vanish. For example the overlap squared in the case of the 
isotropic XY model is $ 2^{-N/2} $ and that for the Ising model in a transverse 
field is
\begin{equation}
 \exp\left\{ {N \over 2\pi} \int^{\pi}_{0} dq\,
     \ln \left({ \epsilon_{q}\!+\!x\!+\!\cos q \over 2\epsilon_{q} }
         \right) \right\} \ .
\label{itf-overlap}
\end{equation}
Where the overlap is non-zero then several possibilities for the asymptotic 
behaviour exist, which do actually arise in the exact solutions of the example 
models -
\begin{itemize}

\item\underline{gapless case}, isotropic XY and critical Ising Model in a 
  transverse Field, Ref.\cite{alm-xy-W,alm-itf-W-98}:\\
  At a critical point, the first excited state gap vanishes and
  \begin{equation}
     \epsilon - \epsilon_0 \sim A |\xi|^{-\gamma} \ ,
  \label{gapless-HW}
  \end{equation}
  as $ \xi \to -\infty $ and if the overlap is finite then 
  $ \Re(\gamma) > 1 $. Therefore the weight function at the
  bottom of the spectrum takes the following form
  \begin{equation}
     w(\epsilon) \sim
     (\epsilon-\epsilon_{0})^{-{1+\gamma \over 2\gamma}}
     \exp\left\{ N {b \over 1\!-\!1/\gamma}
                   (\epsilon-\epsilon_{0})^{1-1/\gamma}
         \right\} \ ,
  \label{gapless-wgt}
  \end{equation}
  This measure is integrable on $ (\epsilon_0,\epsilon_{\infty}) $
  because of the above condition $ \Re(\gamma) > 1 $ and has a branch point 
  at the ground state energy $ \epsilon_{0} $.

\item\underline{gapped case 1}, Ising Model in a transverse Field, in the 
  ordered phase with the disordered trial state, Ref.\cite{alm-itf-W-98}:\\
  if the gap is finite then one possibility is that 
  \begin{equation}
     \epsilon - \epsilon_0 \sim A e^{-\Delta |\xi|} \ ,
  \label{gap1-HW}
  \end{equation}
  as $ \xi \to -\infty $ and where the excited state gap $ \Delta > 0 $.
  One can show that the weight function near the bottom edge of the spectrum 
  is analytic having the form
  \begin{equation}
     w(\epsilon) \sim
     { 1 \over 
       \Gamma(N {[\epsilon-\epsilon_{0}] \over \Delta}\!+\!1) } \ .
  \label{gap1-wgt}
  \end{equation}

\item\underline{gapped case 2}, Ising Model in a transverse Field, in the 
  disordered phase with the disordered trial state, Ref.\cite{alm-itf-W-98}:\\
  and yet another type of gap behaviour exists 
  \begin{equation}
     \epsilon - \epsilon_0 \sim A |\xi|^{-\gamma} e^{-\Delta |\xi|} \ ,
  \label{gap2-HW}
  \end{equation}
  The leading order behaviour of the weight function in this case is
  \begin{equation}
     w(\epsilon) \sim
     (\epsilon-\epsilon_{0})^{-1/2-N(\epsilon-\epsilon_{0})}
     \left[ -\log(\epsilon-\epsilon_{0})
     \right]^{-N\gamma(\epsilon-\epsilon_{0})} \ ,
  \label{gap2-wgt}
  \end{equation}
  which again has a branch point at the bottom edge of the spectrum.

\end{itemize}

So generally we find the support of the measure is bounded which excludes
a number of weight function types such as the Freud or Erd\"os weights, but 
that the weight functions belong to the Szeg\"o class on
$ [\epsilon_0,\epsilon_{\infty}] $,
\begin{equation}
  \int^{\epsilon_{\infty}}_{\epsilon_0} d\epsilon\;
  { \log w(\epsilon) \over 
    \sqrt{[\epsilon_{\infty}-\epsilon][\epsilon-\epsilon_0]} }
  > -\infty \ .
\label{szego}
\end{equation}

\vfill\eject

\stepcounter{section}
\section*{\Roman{section}. Exactly Solvable Lanczos Process}

In this section we derive how the exact Lanczos functions $ \alpha(s) $ and 
$ \beta^2(s) $ can be constructed directly from the knowledge of the connected 
Moments or Cumulants, or more specifically from the Cumulant Generating 
Function.  This is the initial data that one uses in any analysis of quantum 
Many-Body Systems with this approach, and for soluble models the full Generating
Function may be available. However if this is not the case then one would use a 
set of low order Cumulants, up to a given order.

As a first step we recast the Hankel determinants into Selberg Integral form, 
from the classical result\cite{ops-S}
\begin{equation}
    \Delta_n(t) 
    = {1\over (n\!+\!1)!} \int^{+\infty}_{-\infty} 
        \prod^{n+1}_{k=1} d\rho(\epsilon_k) \;
          e^{Nt\sum^{n+1}_{k=1} \epsilon_{k}}
            \prod_{1\leq i<j \leq n+1} |\epsilon_i-\epsilon_j|^2
  \ .
\label{hankel-2}
\end{equation}

For the steps leading to the two conditions which will define the Lanczos 
functions we follow Chen and Ismail\cite{matrix-CI-97}. A similar approach, but 
just confined to the evaluation the Hankel determinants, was taken in References
\cite{ops-BIPZ-78,ops-BBM-92}. The Hankel determinant can be recast into 
the form of a partition function, which is,
\begin{equation}
  \Delta_n(t) = {1 \over (n\!+\!1)!}
            \int^{+\infty}_{-\infty} \prod^{n+1}_{i} d\epsilon_i\;
    \exp\left\{ -\sum^{n+1}_{i} u(\epsilon_i)
                +Nt\sum^{n+1}_{i} \epsilon_i
                +2\sum^{n+1}_{i<j} \ln |\epsilon_{i}-\epsilon_{j}|
        \right\} \ .
\label{hankel-pf}
\end{equation}
One should observe that both $ \sum^{n+1}_{i} u(\epsilon_i) $ and
$ Nt\sum^{n+1}_{i} \epsilon_i $ are of order $ (n\!+\!1)N $ whilst the remaining
term in the argument $ \sum^{n+1}_{i<j} \ln |\epsilon_{i}-\epsilon_{j}| $ is 
of order $ (n\!+\!1)^2 $, so that the only relative scaling that remains
nontrivial is one in which $ n/N $ is fixed. The alternatives would lead to 
completely trivial consequences. The leading order term for this Hankel 
determinant as $ n, N \to \infty $ is given by a steepest descent approximation
(see Ref.~\cite{asymptotic-W} section IX.5)
\begin{equation}
   \Delta_{n}(t) = 
   {(2\pi)^{n+1} \over (n\!+\!1)!} \left| 
   {\partial^2 f \over \partial\epsilon^{0}_{i}\partial\epsilon^{0}_{j}}
                                  \right|^{-1/2}
   e^{-f(\epsilon^{0})}\left[1 + {\rm O}(1/n,1/N)\right] \ ,
\label{hankel-steep}
\end{equation}
where the function $ f(\epsilon) $ is defined as
\begin{equation}
   f(\epsilon) = \sum^{n+1}_{i} u(\epsilon_i)
                -Nt\sum^{n+1}_{i} \epsilon_i
                -2\sum^{n+1}_{i<j} \ln |\epsilon_{i}-\epsilon_{j}| \ ,
\label{saddle-fn}
\end{equation}
and the saddle points $ \{\epsilon^{0}_{i}\}^{n+1}_{i=1} $ are given by
\begin{equation}
   u'(\epsilon^{0}_{i}) = Nt + 2 \sum^{n+1}_{i \neq j} 
   {1 \over \epsilon^{0}_{i}-\epsilon^{0}_{j}} \ .
\end{equation}
\label{saddle-pt}
One can easily show that the Hessian in Eq.~(\ref{hankel-steep}) is positive 
definite given that $ u(\epsilon) $ is convex. One can carry the continuum 
limit further by describing the saddle points as a charged fluid whose dynamics 
are governed by an Energy Functional $ F[\sigma] $
\begin{equation}
  \exp\left( -f(\epsilon^{0}) \right)
  \xrightarrow[n,N \to \infty]{} \exp\left( -F[\sigma_0] \right) \ ,
\label{hankel-fluid}
\end{equation}
with a charge density $ \sigma(\epsilon) $ defined on an interval of 
integration which is to be determined, $ I = (\epsilon_{-},\epsilon_{+}) $. The 
energy functional takes the following form
\begin{equation}
   F[\sigma] = \int_{I} d\epsilon \sigma(\epsilon)
           \left[ u(\epsilon) - Nt \epsilon \right]
           - \int_{I} d\epsilon \int_{I} d\epsilon' 
     \sigma(\epsilon) \ln |\epsilon-\epsilon'| \sigma(\epsilon') \ ,
\label{fluid-functional}
\end{equation}
where the single particle confining potential is controlled by the OPS measure 
and the two-body interaction is a logarithmic type. The result of minimising 
this Functional yields the following singular integral equation for the Charge 
Density 
\begin{equation}
       u'(\epsilon)-Nt = 2\,{\rm PV}\int_{I} d\epsilon'
                { \sigma_{0}(\epsilon') \over \epsilon-\epsilon' } \ .
\label{fluid-singular}
\end{equation}
The solution of this integral equation for the Minimal Charge Density 
$ \sigma_{0}(\epsilon) $ can be found exactly and is
\begin{equation}
       \sigma_{0}(\epsilon) = 
{ \sqrt{ (\epsilon_{+}-\epsilon)(\epsilon-\epsilon_{-}) } \over
  2\pi^2 } \,{\rm PV}\int_{I} d\epsilon'
{ u'(\epsilon')-Nt \over (\epsilon'-\epsilon)
     \sqrt{ (\epsilon_{+}-\epsilon')(\epsilon'-\epsilon_{-}) } } \ .
\label{fluid-soln}
\end{equation}
There are two conditions arising from this solution -
\begin{itemize}
  \item
   the first is a Supplementary Condition which is necessary for the charge 
   density solution to be well defined throughout the interval $ I $
   \begin{equation}
       0 = \int_{I} d\epsilon { u'(\epsilon)-Nt \over 
     \sqrt{ (\epsilon_{+}-\epsilon)(\epsilon-\epsilon_{-}) } } \ ,
   \label{suppl-cond}
   \end{equation}
  \item
   and the Normalisation Condition which simply counts the number of Lanczos 
   steps
   \begin{equation}
       n = { 1 \over 2\pi } \int_{I} d\epsilon\; \epsilon 
   { u'(\epsilon)-Nt \over 
     \sqrt{ (\epsilon_{+}-\epsilon)(\epsilon-\epsilon_{-}) } } \ .
   \label{norm-cond}
   \end{equation}
\end{itemize}
Using this solution for the charge density one can substitute this into the 
original defining equations for the Hankel determinants (the leading order 
approximations) and establish that the Lanczos functions are simply defined by 
the interval $ I $ in this way, $ \epsilon_{\pm} = \alpha \pm 2\beta $.

\begin{theorem}
The Lanczos functions are given implicitly by the two integral equations 
\begin{equation}
       0 = \int^{\alpha+2\beta}_{\alpha-2\beta} 
     d\epsilon { \xi(\epsilon) \over 
     \sqrt{ 4\beta^2 - (\epsilon-\alpha)^2 } } \ ,
\label{suppl-lanczos-eqn}
\end{equation}
\begin{equation}
       s = { 1 \over 2\pi } \int^{\alpha+2\beta}_{\alpha-2\beta}
     d\epsilon { \epsilon \xi(\epsilon) \over 
     \sqrt{ 4\beta^2 - (\epsilon-\alpha)^2 } } \ ,
\label{norm-lanczos-eqn}
\end{equation}
where the model dependent equation for the stationary point $ \xi(\epsilon) $ 
is given by Eq.~(\ref{stationary-2}). 
\end{theorem}
This theorem follows from the previous conditions, namely 
Eqs.~(\ref{suppl-cond},\ref{norm-cond}), and the result for the logarithmic 
derivative of the weight function,
\begin{equation}
  u'(\epsilon) = N \xi(\epsilon) + {\rm O}(\log N) \ .
\label{log-deriv}
\end{equation}
$ \square $

Usually this later equation for the saddle point is also an implicit equation 
and invariably a nonlinear one. In our derivation the scaling $ s=n/N $ remains 
finite whilst $ n,N \to \infty $ emerges naturally and in fact it is difficult 
to see how one could avoid this confluence.

We now give an alternative result for the Lanczos functions which is based on 
the time evolution of the Lanczos $ L $-function.
\begin{theorem}
The Lanczos $ L $-function, in the thermodynamic limit is the solution of the 
following integro-differential equation
\begin{equation}
   L(s,t) = 
   \int^{s}_{0} dr\; r D^2_{t} \log L(s-r,t) + sF^{(2)}(t) \ ,
\label{l-function-tl}
\end{equation}
and the two Lanczos functions are derivable from this via
\begin{equation}
\begin{split}
  \alpha(s) 
  & = \int^{s}_{0} dr\; D_{t}\log L(r,0) + F'(0) \ ,
  \\
  \beta^2(s) 
  & = L(s,0) \ .
\end{split}
\label{coeff-l-function-tl}
\end{equation}
\end{theorem}
The integro-differential equation is simply derived from the discrete 
recurrence, namely Eq.~(\ref{l-function-recur}), after making the observation
that the $ j=n $ term involving $ L_{0}(t) $ has to be separated from the sum 
because it encompasses the initial conditions and is itself not generated by 
the recurrence.$ \square $
 
Finally we give a result equivalent to the theorem above, but which involves 
only scaled forms of the Hankel determinants $ \Delta_{n}(N,t) $ and is the 
differential analogue of the above 
Theorem. 
\begin{definition}
We make the following definition for $ \delta(n,N,t) $ in terms of the Hankel 
Determinant,
\begin{equation}
   \Delta_{n}(N,t) = N^{n(n+1)} 
   \left[ \delta(n,N,t) \right]^{N^2} \ ,
\label{d-defn}
\end{equation}
for $ n \geq 1 $ and $ \Delta_{0}(t) = [ \delta(0,t) ]^N $. 
\end{definition}
\begin{lemma}
The function $ \delta(n,N,t) $ is well defined in the scaling limit
$ n, N \to \infty $. 
\end{lemma}
This follows naturally from the relation of the $ \Delta_{n}(t) $ and the 
Lanczos $ L $-function as given in Eq.~(\ref{D-function}), and the well defined 
scaling of this latter function as demonstrated in the Theorem 3 above.
$ \square $

Then we have the following result -
\begin{theorem}
The Lanczos $ \delta(s,t) $-function satisfies the following partial
differential equation in the thermodynamic limit
\begin{equation}
   \exp\left\{ D^2_{s} \log\delta(s,t) \right\}
  = D^2_{t} \log\delta(s,t) \ ,
\label{d-pde}
\end{equation}
with the boundary condition
\begin{equation}
   \lim_{s \to 0^+} { \log\delta(s,t) \over s } 
   = F(t) \qquad \forall\; t \in {\mathbb R^+} \ ,
\label{d-bc}
\end{equation}
The Lanczos functions are given by
\begin{equation}
\begin{split}
  \alpha(s)  & = \left. D_{t}D_{s} \log\delta(s,t) \right|_{t=0} \ ,
  \\
  \beta^2(s) & = \exp\left\{ D^2_{s} \log\delta(s,0) \right\} \ .
\end{split}
\label{d-lanczos}
\end{equation}
\end{theorem}
Using the scaling relation above, Eq.~(\ref{d-defn}), and the equation of motion
for $ \Delta_{n}(t) $, Eq.~(\ref{D-function-eqn}), the result follows.
$ \square $

These last two theorems relate to the dynamics of a nonlinear continuum Toda 
Lattice in one space domain $ s \in {\mathbb R^+} $ and one time domain $ t $, 
with boundary conditions defined at the origin $ s = 0 $ for all times $ t $ 
by the cumulant generating function $ F(t) $. The object is then to find the 
Lanczos functions $ \alpha(s), \beta^2(s) $ from a solution of this system, 
wherein these functions are directly related to the solution at a given time 
$ t = 0 $ over all spatial points $ s > 0 $.

\vfill\eject

\stepcounter{section}
\section*{\Roman{section}. The Taylor Series Expansion}

The investigation of the Taylor series expansion of the Lanczos coefficients 
about $ s=0 $, is an essential element in the application of this Lanczos 
method, as was indicated earlier, where one has only a finite set of low order
cumulants available, say for non-integrable models. Therefore in this case one 
can only construct a truncated Taylor series expansion and so issues concerning 
convergence, the radius of convergence of the series, and whether one can 
extrapolate immediately arise. In addition one would like a direct algorithm 
relating the cumulants to the Lanczos functions from a purely practical point of
view.

We define the Taylor series expansion of the two Lanczos functions by two new 
sequences of coefficients,
\begin{equation}
\begin{split}
  \alpha(s) & = c_1 + \sum^{\infty}_{n=0} a_{n}s^{n+1} \ ,
            \\
  \beta^2(s) & = \sum^{\infty}_{n=0} b_{n}s^{n+1} \ .
\end{split}
\label{coeff-taylor-exp}
\end{equation}
In order to find these coefficients one could use either of the two general 
solutions for the Lanczos process, 
Eqs.~(\ref{suppl-lanczos-eqn},\ref{norm-lanczos-eqn}) or 
Eq.~(\ref{l-function-tl}), and the two methods are presented below.

The first step involves the inversion of the following Taylor series expansion
\begin{equation}
  \epsilon = c_{1} + \sum_{n=1} {c_{n+1} \over n!} \xi^{n} \ ,
\label{pe-energy}
\end{equation}
for $ \xi(\epsilon) $, namely the coefficients $ e_{k} $ appearing in
\label{pe-xi}
\begin{equation}
  \xi = \sum_{k=1} e_{k} (\epsilon-c_{1})^k \ .
\end{equation}
The coefficients $ c_{n} $ appearing in Eq.~\ref{pe-energy} are the cumulant 
coefficients. The existence of this inverse function is guaranteed because the 
second cumulant $ c_2 > 0 $ in all systems and we assume that the saddle point 
function, Eq.~(\ref{stationary-2}), is analytic in the neighbourhood of 
$ \xi = 0 $. The next step involves the solution of the two recurrences 
\begin{equation}
 \begin{split}
   0 & =
       \sum_{k=1} e_{k} \sum^{\lfloor k/2\rfloor}_{m=0}
       \binom{k}{2m} {(1/2)_{m} \over m!} 
       (\alpha-c_{1})^{k-2m} (4\beta^2)^{m} \ ,
     \\
  2s & = 
       \sum_{k=1} e_{k} \sum^{\lfloor (k-1)/2\rfloor}_{m=0}
       \binom{k}{2m\!+\!1} {(1/2)_{m+1} \over (m\!+\!1)!} 
       (\alpha-c_{1})^{k-2m-1} (4\beta^2)^{m+1} \ ,
 \end{split}
\label{pe-L}
\end{equation}
which are used to solve for the coefficients $ a_{n}, b_{n} $ appearing in 
Eq.~(\ref{coeff-taylor-exp}).

In the second method we define a continuum version of the coefficients that are 
defined in Eq.~(\ref{l-function-coeff-1}) in the following way
\begin{equation}
  \log {L(s,t) \over sl_{1}(t)} 
  = \log\left( 1+\sum_{p \geq 1} { l_{p+1} \over l_{1}} s^{p} \right)
  \equiv \sum_{p \geq 1} { m_{p}(t) s^{p} } \ ,
\label{l-function-taylor}
\end{equation}
and the inverse of Eq.~(\ref{l-function-coeff-2}) in an explicit form
\begin{equation}
  {l_{p+1} \over l_{1}} = \sum_{\sum_{i} q_{i}r_{i} = p}
             \prod_{i} {1\over q_{i}!} m_{r_{i}}^{q_{i}} \ .
\label{l-function-coeff-inv}
\end{equation}
From these relations one can find a hierarchy of equations for these 
coefficients
\begin{equation}
\begin{split}
  l_{1}(t) 
  & = F''(t) \ ,
  \\
  l_{2}(t)
  & = { F^{(2)}F^{(4)} - (F^{(3)})^2 \over 2(F^{(2)})^2 } \ ,
  \\
  l_{p+2}(t)
  & = {m''_{p}(t) \over (p\!+\!2)(p\!+\!1)}
    = l_{1}(t) \sum_{\sum_{i} q_{i}r_{i} = p+1}
               \prod_{i} {m_{r_{i}}^{q_{i}} \over q_{i}!}
  \qquad \text{for } p \geq 1 \ .
\end{split}
\label{l-function-taylor-recur}
\end{equation}

Thus one can verify from the solution for the initial value problem above that 
the general Taylor series coefficients are given by
\begin{equation}
\begin{split}
   [(n\!+\!1)!]^2 c^{3n\!+\!1}_{2} a_n
  &  =
   \sum_{\lambda \vdash 2n+1} A(n;\lambda)
   \prod^{2n+1}_{i=0} c^{a_i}_{2+i}
  \\
   (n\!+\!1)!n! c^{3n\!-\!1}_{2} b_n
  &  =
   \sum_{\lambda \vdash 2n} B(n;\lambda)
   \prod^{2n}_{i=0} c^{a_i}_{2+i}
  \ ,
\end{split}
\label{taylor-lanczos}
\end{equation}
where the coefficients labeled by the partition 
$ \lambda = (1^{a_1}.2^{a_2} \ldots i^{a_i}) $, denoted by
$ A(n;\lambda), B(n;\lambda) $, are listed in Table(\ref{lanczos-table}) of 
the Appendix. There are constraints operating in the above equations, namely
$ \sum^{2n+1}_{i=1} ia_i = \sum^{2n+1}_{i=0} a_i = 2n\!+\!1 $ for the first 
relation and $ \sum^{2n}_{i=1} ia_i = \sum^{2n}_{i=0} a_i = 2n $ for the second.

Clearly the Taylor series expansion of the Lanczos functions has low order 
coefficients which are constructed from the low order cumulants, and is a form 
of a linked cluster expansion. However it is not just a simple linked cluster 
expansion as in the Taylor series expansion of the Cumulant Generating Function,
but involves a subtle interplay and cancellation of all cumulants below a given 
order.
  
\vfill\eject

\stepcounter{section}
\section*{\Roman{section}. General Properties}

There are some very general properties that the Lanczos process in the 
thermodynamic limit and the associated Lanczos functions satisfy and we examine 
these now. Some are quite obvious and not particularly surprising, however we 
state these for completeness sake, while there are some other properties which 
are not so immediate but very important nevertheless.

The next, and natural, property concerns the monotonicity of the two envelope 
functions $ \epsilon_{\pm}(s) = \alpha(s) \pm 2\beta(s) $. 
\begin{theorem}
The envelope functions $ \epsilon_{+}(s), \epsilon_{-}(s) $ are monotonically 
increasing and decreasing functions of real, positive $ s $ respectively.
\end{theorem}
This follows from a recasting of the normalisation condition in the following 
way
\begin{equation}
  2\pi s = \int^{\xi{+}}_{\xi_{-}} d\xi 
   \sqrt{ [\epsilon(\xi_{+})-\epsilon(\xi)]
          [\epsilon(\xi)-\epsilon(\xi_{-})] } \ ,
\label{norm-lanczos-2}
\end{equation}
where the $ \xi_{\pm} $ are defined by $ \epsilon(\xi_{\pm}) = \epsilon_{\pm} $.
Now it is straight forward to write the explicit forms for the derivatives of 
the envelope functions with respect to $ s $ as
\begin{equation}
 \begin{split}
   {d \epsilon_{+} \over ds} & = 4\pi \Big/
   \int^{\xi_{+}}_{\xi_{-}} d\xi 
   \sqrt{ \epsilon(\xi)-\epsilon(\xi_{-}) \over
          \epsilon(\xi_{+})-\epsilon(\xi) } \ ,
   \\       
   {d \epsilon_{-} \over ds} & = -4\pi \Big/
   \int^{\xi_{+}}_{\xi_{-}} d\xi 
   \sqrt{ \epsilon(\xi_{+})-\epsilon(\xi) \over
          \epsilon(\xi)-\epsilon(\xi_{-}) }
   \ ,
 \end{split}
\label{lanczos-mono}
\end{equation}
so that the stated properties are evident.$ \square $

It is clear that the envelope functions $ e_{\pm}(s) $ are bounded in the 
following ways, $ \epsilon_{-}(s) \geq \epsilon_{0} $ and 
$ \epsilon_{+}(s) \leq \epsilon_{\infty} $.

The 3-term recurrence which serves as one of the definitions of the Orthogonal 
Polynomials themselves is now going to take a definite limiting form when
$ n, N \to \infty $ such that $ s $ is finite. This is going to lead to a 
scaling form for one set of the Polynomials themselves, which would be more 
correctly termed orthogonal functions $ p(s,\epsilon) $. Heuristically one can 
see how this arises by the following argument. If one ensures that Lanczos 
densities are employed and the following scaling of the polynomials thus
$ P_n(E) = N^n p_n(\epsilon) $, then the 3-term recurrence becomes
\begin{equation}
  p_{n+1}(\epsilon)/p_n(\epsilon) +
  \beta^2_n {1 \over p_{n}(\epsilon)/p_{n-1}(\epsilon)} =
  \epsilon-\alpha_n \ .
\label{ops-3-term}
\end{equation}
Now these ratios are approximated by
\begin{equation}
  {p_{n+1}(\epsilon) \over p_n(\epsilon)} \sim
 \exp\left( {1 \over N}{\partial \over \partial s} \ln p(s,\epsilon)
     \right) \ ,
\label{ops-ratio}
\end{equation}
for arguments $ \epsilon \in {\mathbb C}\backslash{\rm Supp}[d\rho] $. So that 
in the asymptotic regime the recurrence becomes
\begin{equation}
 \exp\left( {1 \over N}{\partial \over \partial s} \ln p(s,\epsilon) \right)
  + \beta^2(s)
 \exp\left(-{1 \over N}{\partial \over \partial s} \ln p(s,\epsilon) \right)
  \sim \epsilon-\alpha(s) \ ,
\label{ops-recur-tl}
\end{equation}
whose solutions are
\begin{equation}
  p^{\pm}(s,\epsilon) \sim p(0)
  \exp\left\{ N\int^s dt \ln \half
      \left[ \epsilon-\alpha(t) \pm 
             \sqrt{ (\epsilon-\alpha(t))^2 - 4\beta^2(t) } \right]
      \right\} \ .
\label{poly-tl}
\end{equation}
These are the corresponding results for the ratio $ P_{n}(x)/P_{n+1}(x) $ or 
n-th root $ \sqrt[n]{P_{n}(x)} $ asymptotics of generic Orthogonal Polynomials 
as $ n \to \infty $\cite{ops-S,ops-F,ops-N,ops-N-79,ops-vA}, or the scaled 
Orthogonal Polynomials\cite{ops-vA-90}, but are rather different due to the 
particular nature of Many-Body Orthogonal Polynomials.
\begin{theorem}
Given the scaling behaviour of the Lanczos coefficients, and that they are 
bounded for $ n, N \to \infty $, then the n-th root of the denominator 
Orthogonal Polynomials $ p_{n}(\epsilon) $ have the limiting form uniformly 
for $ \epsilon $ in compact subsets of 
$ {\mathbb C}\backslash{\rm Supp}[d\rho] $.
\begin{equation}
  p(s,\epsilon) \equiv 
  \lim_{n, N \to \infty}|p_{n}(N,\epsilon)|^{1/N} =
  \exp\left\{ \int^{s}_{0} dt\; \ln \half
      \left[ \epsilon\!-\!\alpha(t) +
             \sqrt{ (\epsilon\!-\!\alpha(t))^2\! -\! 4\beta^2(t) } \right]
      \right\}
\label{poly-limit}
\end{equation}
\end{theorem}
The proof of this parallels the one constructed by van Assche in 
Ref.~\cite{ops-vA-90} through the use of Tur\'an Determinants, 
\begin{equation}
  D_{n} \equiv p^{2}_{n} -  p_{n+1}p_{n-1}
\label{turan-D}
\end{equation}
One can show that these obey the following recurrence relation
\begin{equation}
  D_{n} = \beta^2_{n}D_{n-1}
          + (\alpha_{n}\!-\!\alpha_{n-1})p_{n}p_{n-1}
          + (\beta^2_{n}\!-\!\beta^2_{n-1})p_{n}p_{n-2}
\label{turan-recur}
\end{equation}
Using the partial fraction decomposition of the ratio of two successive
Orthogonal Polynomials one can also find a bound on this ratio
\begin{equation}
  \left|{ p_{n-1}(\epsilon) \over p_{n}(\epsilon) }\right|
  \leq {C \over d} \qquad \forall\; n
\label{ratio-bound}
\end{equation}
for all $ \epsilon \in K $ where the compact set 
$ K \subset {\mathbb C}\backslash{\rm Supp}[d\rho] $ and $ d $ is the 
distance between this set and the interval 
$ [\epsilon_{0},\epsilon_{\infty}] $, and $ C $ is a positive constant.
Using Eq.~(\ref{turan-recur}) we have
\begin{equation}
  \left|{ D_{n} \over p^2_{n} }\right| \leq
    \sup_{n}(\beta^2_{n}){C^2 \over d^2}
    \left|{ D_{n-1} \over p^2_{n-1} }\right|
      + |\alpha_{n}-\alpha_{n-1}|{C \over d}
        + |\beta^2_{n}-\beta^2_{n-1}|{C^2 \over d^2}
\label{turan-bound}
\end{equation}
Given the scaling form of the Lanczos coefficients the ratio
$ |D_{n}/ p^2_{n}| \to 0 $ as $ n, N \to \infty $ uniformly in $ \epsilon $ 
whenever $ d $ is large enough. This means that $ |p_{n-1}/p_{n}| $ and 
$ |p_{n}/p_{n+1}| $ tend to the same accumulation point which we denote by 
$ p(s,\epsilon) $. This point is given by the solution of the quadratic 
equation $ p + \beta^2(s)/p = \epsilon-\alpha(s) $, and the positive branch of 
the solution must be taken as $ p \to \infty $ when $ \epsilon \to \infty $. 
The functions $ p(s,\epsilon) $ are analytic functions of $ \epsilon \in K $ 
which are uniformly bounded, so the restriction on $ d $ can be lifted to being 
only non-zero. The behaviour of the n-th ratio then gives the n-th root 
behaviour directly as 
\begin{equation}
  \left| p_{n} \right|^{1/N} =
  \exp\left\{ {1\over N} \sum_{k=1}^{n}
              \log \left|{ p_{k}(\epsilon) \over p_{k-1}(\epsilon) }
                   \right|
      \right\}
\label{root-ratio}
\end{equation}
The asymptotic behaviour that we have found applies to the denominator OP only 
as can be seen from the observation that $ p_{1} = \epsilon-c_{1} $ and 
$ p_{2} = (\epsilon-c_{1})^2 - c_{3}/c_{2}N(\epsilon-c_{1}) - c_{2}/N $, while 
\begin{equation}
   \left[ \epsilon\!-\!\alpha(s) +
             \sqrt{ (\epsilon\!-\!\alpha(s))^2\! -\! 4\beta^2(s) } \right]
   \xrightarrow[s \to 0]{}
   {1\over \epsilon-c_{1}}
   \left( (\epsilon-c_{1})^2 - c_{3}/c_{2}N(\epsilon-c_{1}) - c_{2}/N 
   \right) \ .
\label{initial-OP}
\end{equation}
This establishes the result.$\square$

\vfill\eject

\stepcounter{section}
\section*{\Roman{section}. Summary}

In this work we have demonstrated the general scaling behaviour of the Lanczos
Process as applied to Many-Body Systems when the process is taken to 
convergence and the thermodynamic limit taken. We also find explicit 
constructions of the limiting Lanczos coefficients in two equivalent 
formulations, from an initial exact solution of the moment problem, that is to 
say the cumulant generating function for the system. There are explicit 
examples where the CGF can be found and the whole Lanczos process explicitly
realised. Furthermore we have given the corresponding results for the associated
Orthogonal Polynomial system and the measure in this regime, quite generally. 
However we must emphasise that these results apply only to the bulk properties,
that is to say the ground state properties that scale extensively and the 
spectral properties in the interior (the "bulk") of the spectrum. So this does
not include the delicate scaling behaviour at the edges of the spectrum, nor in
the neighbourhood of singularities - this theory would have to be extended to
treat the excited state gaps near the bottom of the spectrum. A number of 
general theorems are given which constrain the behaviour of the Lanczos 
functions, and the process in general. We also indicate how a number of such 
constraints operating can lead to some concrete realisations or scenarios that 
the Lanczos process can present, namely its behaviour at a critical point in 
the model under study. This is a significant step on the way to the goal of a 
rigorous classification of Many-Body Systems in terms of their character via 
the Lanczos process. Other important questions that arise in the treatment of
non-integrable models, for which the general results presented here have
suggested some answers, are the questions of the choice of trial state, the 
rate of convergence of the truncated Lanczos process and how one might 
accelerate its convergence given some independent qualitative knowledge.

\vfill\eject

\bibliographystyle{jmp}
\bibliography{moment,texp,cmx,pexp,alm,math}

\newcommand{\noopsort}[1]{} \newcommand{\printfirst}[2]{#1}
  \newcommand{\singleletter}[1]{#1} \newcommand{\switchargs}[2]{#2#1}
\begin{thebibliography}{10}
\newcommand{\enquote}[1]{``#1''}

\bibitem{lanczos-D-94}
E.~Dagotto,
\newblock \enquote{Correlated Electrons in high-temperature Superconductors.}
\newblock Rev. Mod. Phys. 66, 763--840 (1994)

\bibitem{lanczos}
C.~Lanczos,
\newblock J. Res. Natl. Bur. Stand. 45, 255 (1950)

\bibitem{eigen-P}
B.~N. Parlett,
\newblock {\em The Symmetric Eigenvalue Problem\/}.
\newblock Prentice-Hall, Englewood Cliffs (1980)

\bibitem{eigen-C}
F.~Chatelin,
\newblock {\em Eigenvalues of Matrices\/}.
\newblock Wiley, Chichester and New York (1993)

\bibitem{eigen-S}
Y.~Saad,
\newblock {\em Numerical Methods for Large Eigenvalue Problems\/}.
\newblock Manchester University Press Series in Algorithms and Architectures
  for Advanced Scientific Computing (1991)

\bibitem{leb-K}
S.~Kaniel,
\newblock \enquote{Estimates for some Computational Techniques in Linear
  Algebra.}
\newblock Math. Comp. 20, 369--378 (1966)

\bibitem{leb-P}
C.~C. Paige,
\newblock \enquote{Error Analysis of the Lanczos Algorithm for tridiagonalizing
  a symmetric Matrix.}
\newblock J. Inst. Math. Appl. 18, 341--349 (1976)

\bibitem{leb-S}
Y.~Saad,
\newblock \enquote{On the Rates of Convergence of the Lanczos and the
  Block-Lanczos Methods.}
\newblock SIAM J. Numer. Anal. 17, 687--706 (1980)

\bibitem{ops-S}
G.~Szeg{\"o},
\newblock {\em Orthogonal Polynomials\/}.
\newblock Colloquium Publications {\bf 23}. American Mathematical Society,
  Providence, Rhode Island, 4th edn. (1975)

\bibitem{ops-C}
T.~S. Chihara,
\newblock {\em An Introduction to Orthogonal Polynomials\/}.
\newblock Gordon and Breach, New York (1978)

\bibitem{ops-F}
G.~Freud,
\newblock {\em Orthogonal Polynomials\/}.
\newblock Pergamon Press, Oxford (1971)

\bibitem{mm-L89a}
B.~G. Lindsay,
\newblock \enquote{On the Determinants of Moment Matrices.}
\newblock Ann. Stat. 17, 711--721 (1989)

\bibitem{matrix-CI-97}
Y.~Chen and M.~E.~H. Ismail,
\newblock \enquote{Thermodynamic relations of the Hermitian matrix ensembles.}
\newblock J. Phys. A: Math. Gen. 30, 6633--6654 (1997)

\bibitem{ops-vA-90}
W.~van Assche,
\newblock \enquote{Asymptotics for Orthogonal Polynomials and three-term
  Recurrences.}
\newblock In P.~Nevai, editor, {\em Orthogonal Polynomials\/}, pp. 435--462.
  Kluwer Academic, Dordrecht (1990)

\bibitem{moment-ST}
J.~A. Shohat and J.~D. Tamarkin,
\newblock {\em The Problem of Moments\/}.
\newblock American Mathematical Society, Providence, Rhode Island (1943)

\bibitem{moment-A}
N.~I. Akhiezer,
\newblock {\em The Classical Moment Problem\/}.
\newblock Oliver and Boyd, London (1965)

\bibitem{cf-JT}
W.~B. Jones and W.~J. Thron,
\newblock {\em Continued Fractions - Analytic Theory and Applications\/}.
\newblock Addison-Wesley Publishing Company, Reading, Massachusetts (1980)

\bibitem{cf-LW}
L.~Lorentzen and H.~Waadeland,
\newblock {\em Continued Fractions with Applications\/}.
\newblock North-Holland, Amsterdam (1992)

\bibitem{moment-K}
M.~G. Kendall,
\newblock {\em The Advanced Theory of Statistics\/}, vol.~1.
\newblock Edward Arnold and Halstead Press, London and New York, 6th edn.
  (1994)

\bibitem{cumulant-K}
R.~Kubo,
\newblock \enquote{Generalized Cumulant Expansion Method.}
\newblock J. Phys. Soc. Japan 17, 1100--1120 (1962)

\bibitem{pexp-tgap-HWW}
L.~C.~L. Hollenberg, M.~P. Wilson, and N.~S. Witte,
\newblock \enquote{General non-perturbative massgap to first order in 1/V.}
\newblock Phys. Lett. B 361, 81--86 (1995)

\bibitem{texp-W-97}
N.~S. Witte,
\newblock \enquote{Analytic Solution to the Moment Problem for the XY Chain.}
\newblock Int. J. Mod. Phys. B 11, 1503--1517 (1997)

\bibitem{alm-itf-W-98}
N.~S. Witte,
\newblock \enquote{Moment Formalisms applied to a solvable Model with a Quantum
  Phase Transition. II. Geometrical Moment Methods.}
\newblock submitted to Nuc. Phys. B  (1999)

\bibitem{positive-K}
S.~Karlin,
\newblock {\em Total Positivity\/}, vol.~1.
\newblock Stanford University Press, Stanford (1968)

\bibitem{ops-T}
M.~Toda,
\newblock {\em Theory of Nonlinear Lattices\/}.
\newblock Springer Series in Solid-State Sciences 20. Springer-Verlag, Berlin,
  2nd edn. (1989)

\bibitem{pexp-proof-WH}
N.~S. Witte and L.~C.~L. Hollenberg,
\newblock \enquote{Plaquette Expansion Proof and Interpretation.}
\newblock Z. Phys. B 95, 531--539 (1994)

\bibitem{pexp-1st-H}
L.~C.~L. Hollenberg,
\newblock \enquote{Plaquette Expansion in lattice Hamiltonian Models.}
\newblock Phys. Rev. D 47, 1640--1644 (1993)

\bibitem{pexp-exactgse-HW}
L.~C.~L. Hollenberg and N.~S. Witte,
\newblock \enquote{Analytic Solution for the Ground State Energy of the
  Extensive Many-Body Problem.}
\newblock Phys. Rev. B 54, 16309--16312 (1996)

\bibitem{pexp-2dhm-WHW}
N.~S. Witte, L.~C.~L. Hollenberg, and Z.~Weihong,
\newblock \enquote{Two-dimensional XXZ Model ground state Properties using an
  analytic Lanczos Expansion.}
\newblock Phys. Rev. B 55, 10412--10419 (1997)

\bibitem{asymptotic-W}
R.~Wong,
\newblock {\em Asymptotic Approximations of Integrals\/}.
\newblock Academic Press, Boston (1989)

\bibitem{cmx-overlap-C}
J.~Cioslowski,
\newblock \enquote{Estimation of the overlap between the approximate and exact
  wave function of the ground state from the connected-moments expansion.}
\newblock Phys. Rev. A 36, 3441--3442 (1987)

\bibitem{alm-xy-W}
N.~S. Witte,
\newblock \enquote{The exact realisation of the Lanczos Method for a quantum
  Many-Body System.}
\newblock Phys. Lett. A 254, 18--23 (1999)

\bibitem{ops-BIPZ-78}
E.~Br{\'e}zin, C.~Itzykson, G.~Parisi, and J.~B. Zuber,
\newblock \enquote{Planar Diagrams.}
\newblock Commun. Math. Phys. 59, 35--51 (1978)

\bibitem{ops-BBM-92}
G.~{Baker Jr.}, D.~Bessis, and P.~Moussa,
\newblock \enquote{Asymptotic Behaviour of some Hankel-Toeplitz Determinants.}
\newblock Rev. Math. Phys. 4, 65--94 (1992)

\bibitem{ops-N}
P.~G. Nevai,
\newblock {\em Orthogonal Polynomials\/}.
\newblock {\bf }. American Mathematical Society, Providence, Rhode Island

\bibitem{ops-N-79}
P.~G. Nevai,
\newblock \enquote{Distribution of Zeros of Orthogonal Polynomials.}
\newblock Trans. Amer. Math. Soc. 249, 341--361 (1979)

\bibitem{ops-vA}
W.~van Assche,
\newblock {\em Asymptotics of Orthogonal Polynomials\/}.
\newblock Lecture Notes in Mathematics {\bf 23}. Springer Verlag, Berlin

\end{thebibliography}
\vspace*{30mm}
\begin{center}
\bf Acknowledgements
\end{center}
One of the authors (NSW) would like to acknowledge the support of a
Australian Research Council large Grant whilst this work was performed, and
the hospitality of Service de Physique Th\'eorique, Centre d'Etudes 
Nucl\'eaires de Saclay.
\vfill\eject

\section*{Appendix}

We list here the coefficients of the Taylor series expansion for the Lanczos
Coefficients, labelled by the partitions of integers, according to the 
definition of Eq.(\ref{taylor-lanczos}).

\renewcommand{\arraystretch}{1.2}
\begin{table}

\begin{tabular}{|cc|c|r|}
\hline
 $ 1 $ & $ a_0 $ & $ \lambda = $ & $ A(0;\lambda) = $ \\
\cline{3-4}
&
& $ 1 $
& $ 1 $ \\
\hline

 $ 2 $ & $ b_1 $ & $ \lambda = $ & $ B(1;\lambda) = $ \\
\cline{3-4}
&
& $ 2 $
& $  1 $ \\
&
& $ 1^2 $
& $ -1 $ \\
\hline

 $ 3 $ & $ a_1 $ & $ \lambda = $ & $ A(1;\lambda) = $ \\
\cline{3-4}
&
& $ 1^3 $
& $ 3 $ \\
&
& $ 2.1 $
& $ -4 $ \\
&
& $ 3 $
& $ 1  $ \\
\hline

 $ 4 $ & $ b_2 $ & $ \lambda = $ & $ B(2;\lambda) = $ \\
\cline{3-4}
&
& $ 1^4 $
& $ -12 $ \\
&
& $ 2.1^2 $
& $ 21 $ \\
&
& $ 2^2 $
& $ -4 $ \\
&
& $ 3.1 $
& $ -6 $ \\
&
& $ 4 $
& $ 1 $ \\
\hline

 $ 5 $ & $ a_2 $ & $ \lambda = $ & $ A(2;\lambda) = $ \\
\cline{3-4}
&
& $ 1^5 $
& $ 81 $ \\
&
& $ 2.1^3 $
& $ -174 $ \\
&
& $ 3.1^2 $
& $ 48 $ \\
&
& $ 2^2.1 $
& $ 70 $ \\
&
& $ 4.1 $
& $ -9 $ \\
&
& $ 3.2 $
& $ -17 $ \\
&
& $ 5 $
& $  1  $ \\
\hline

 $ 6 $ & $ b_3 $ & $ \lambda = $ & $ B(3;\lambda) = $ \\
\cline{3-4}
&
& $ 1^6 $
& $ -567 $ \\
&
& $ 2.1^4 $
& $ 1449 $ \\
&
& $ 3.1^3 $
& $ -414 $ \\
&
& $ 2^2.1^2 $
& $ -872 $ \\
&
& $ 4.1^2 $
& $ 84 $ \\
&
& $ 3.2.1 $
& $ 304 $ \\
&
& $ 5.1 $
& $ -12 $ \\
&
& $ 2^3 $
& $ 70 $ \\
&
& $ 4.2 $
& $ -26 $ \\
&
& $ 3^2 $
& $ -17 $ \\
&
& $ 6 $
& $ 1   $ \\
\hline

\end{tabular}
\begin{tabular}{|cc|c|r|}
\hline
 $ 7 $ & $ a_3 $ & $ \lambda = $ & $ A(3;\lambda) = $ \\
\cline{3-4}
&
& $ 1^7 $
& $ 5805$ \\
&
& $ 2.1^5 $
& $ -17190$ \\
&
& $ 3.1^4 $
& $ 4815$ \\
&
& $ 2^2.1^3 $
& $ 13940$ \\
&
& $ 4.1^3 $
& $ -990$ \\
&
& $ 5.1^2 $
& $ 150$ \\
&
& $ 3.2.1^2 $
& $ -5470$ \\
&
& $ 2^3.1 $
& $ -2680$ \\
&
& $ 3^2.1 $
& $ 425$ \\
&
& $ 4.2.1 $
& $ 680$ \\
&
& $ 6.1 $
& $ -16$ \\
&
& $ 3.2^2 $
& $ 640$ \\
&
& $ 5.2 $
& $ -44$ \\
&
& $ 4.3 $
& $ -66$ \\
&
& $ 7 $
& $ 1 $ \\
\hline

 $ 8 $ & $ b_4 $ & $ \lambda = $ & $ B(4;\lambda) = $ \\
\cline{3-4}
&
& $ 1^8 $
& $ -58050 $ \\
&
& $ 2.1^6 $
& $ 195345 $ \\
&
& $ 3.1^5 $
& $ -55710 $ \\
&
& $ 2^2.1^4 $
& $ -197470 $ \\
&
& $ 3.2.1^3 $
& $ 85430 $ \\
&
& $ 4.1^4 $
& $ 11745 $ \\
&
& $ 5.1^3 $
& $ -1890 $ \\
&
& $ 2^3.1^2 $
& $ 60580 $ \\
&
& $ 3^2.1^2 $
& $ -8020 $ \\
&
& $ 4.2.1^2 $
& $ -12520 $ \\
&
& $ 6.1^2 $
& $ 230 $ \\
&
& $ 3.2^2.1 $
& $ -22820 $ \\
&
& $ 4.3.1 $
& $ 1860 $ \\
&
& $ 5.2.1 $
& $ 1200 $ \\
&
& $ 7.1 $
& $ -20 $ \\
&
& $ 2^4 $
& $ -2680 $ \\
&
& $ 4.2^2 $
& $ 1320 $ \\
&
& $ 3^2.2 $
& $ 1705 $ \\
&
& $ 6.2 $
& $ -60 $ \\
&
& $ 5.3 $
& $ -110 $ \\
&
& $ 4^2 $
& $ -66 $ \\
&
& $ 8 $
& $ 1 $ \\
\hline

\end{tabular}

\bigskip
\caption{
The coefficients in the Taylor series expansion for the Lanczos functions
$ \alpha(s) $ and $ \beta^2(s) $, as defined in Eq.(\ref{taylor-lanczos}),
and the labels denoting the partitions $ \lambda $ of the positive integers.
}
\label{lanczos-table}
\bigskip

\end{table}

\end{document}